# Offloading traffic hotspots using moving small cells

Aymen Jaziri *+, Ridha Nasri *, Tijani Chahed +
*Orange Labs, +Telecom Sudparis
Email: {aymen.jaziri, ridha.nasri}@orange.com, tijani.chahed@telecom-sudparis.eu

*Abstract*—In this paper, the concept of moving small cells in mobile networks is presented and evaluated taking into account the dynamics of the system. We consider a small cell moving according to a Manhattan mobility model which is the case when the small cell is deployed on the top of a bus following a predefined trajectory in areas which are generally crowded. Taking into account the distribution of user locations, we study the dynamic level considering a queuing model composed of multi-class Processor Sharing queues. Macro and small cells are assumed to be operating in the same bandwidth. Consequently, they are coupled due to the mutual interferences generated by each cell to the other. Our results show that deploying moving small cells could be an efficient solution to offload traffic hotspots.

## I. INTRODUCTION

With the exponential growth of data traffic in modern mobile networks and the emergence of variety of connected devices, the presence of traffic hotspots, reflecting the occurrence of mass events or the existence of areas of capacity bottlenecks, has become one of important scenarios to take into consideration in operators roadmaps to reach their objectives in terms of quality of service (QoS). Therefore, analysis of small cell deployment under the presence of traffic hotspots is a relevant issue to investigate and also to include in network planning process, as it allows to evaluate the efficiency of HetNet deployments. Moreover, considering the mobility of small cells is a relevant subject which falls in the area of network densification which is among the dominant themes for wireless evolution into 5G networks [1].

The efficiency of small cell deployment was the subject matter of several works [2]–[5]. Authors in [2] considered a HetNet with different tiers to evaluate the average achievable rate and the average load per tier. The network structure in each tier is based on a spatial Poisson Point Process (PPP). A different approach was used in [4] where the authors used a fluid model in order to study the impact of small cell location on the performance of HetNets. Performance analysis of cellular networks incorporating sophisticated queuing models have been also well investigated from a dynamic level point of view [5]. In reference [5], authors showed that the densification of the network with small cells is globally a good solution to offload traffic. Authors in [6], [7] provided static and dynamic level analysis of small cell deployment under imperfect hotspot localization and results in [6], [7] showed that small cells do not always generate positive capacity gains.

In contrary to the deployment of classic HetNets, the idea of moving small cells is a fresh topic not exhaustively investigated in the literature. In [9], authors presented the benefits and the challenges of using moving relay nodes and applied a simulation approach in order to evaluate the capacity gains generated from deploying this technology.

In the same context, we study a scenario where the small cell is moving according to Manhattan mobility model (it models a movement in city streets environment such as the classic bus trajectory using grid road topology) and we consider a traffic hotspot (reflecting a mass event such as a street show). We derive the throughput distribution and incorporate it as an input in a dynamic level evaluation based on a network of coupled multi-class PS queues with elastic traffic. The coupling between the macro cell and the small cell is the consequence of the interference generated by each cell to the other one when at least one user is served by the interfering cell. From a practical point of view, the objective of this study is to know if the system generates positive offloading gains from deploying moving small cells and also to understand when the deployment of moving small cells is worthless. So far, this evaluation allows operators investing in an efficient way to retain the appropriate solution in offloading traffic.

The remainder of this paper is organized as follows. In section II, we present the adopted concept of moving small cells and we detail the considered mobility model for them. In section III, we describe the downlink system model with a special focus on radio aspects. In section IV, a preliminary study is presented allowing to derive several inputs of the dynamic level analysis. Section V details the analysis for the above-mentioned scenario. Numerical results are highlighted in section VI. In section VII, we conclude the paper and indicate some directions for future works.

## II. MOVING CELLS AND MANHATTAN MOBILITY MODEL

### A. Moving small cells concept and evolution

Moving small cell is a new concept and a fresh topic which could be a relevant solution to offload traffic. It is still not actively discussed in research and industrial communities because until recently, they are trying to boost network performances with lower operational costs. Mobility of small cells is the most advantageous feature allowing to efficiently offload moving and/or unpredictable congesting traffic leading thus to improve the network performances. In this work, we suppose that the small cell is deployed on the top of public transportation means which are generally circulating in very crowded streets. For instance, on the top of a bus or a taxi, a small cell allows to carry traffic generated by its own passengers in addition to the one coming from UEs (User Equipment) in its vicinity. Consequently, the data rate of each user in the coverage area will be improved since the line-of-sight is guaranteed and the effect of slow fading is reduced. In the presence of a stationary traffic hotspot, it is clear that operators do not need moving small cells. Nevertheless, if this traffic hotspot is temporary, then deploying classic small cells would be costly and inefficient (even with the cell shutdown concept because not all the energy-consuming components in the base station are turned off in case of small load.) and it turns out credible to replace them by smaller number of moving small cells. Consequently, studying the scenario of moving small cells allows to evaluate the possible capacity and QoS gains in case the traffic hotspot is not moving or is following a different trajectory compared to the small cell one. The results allow operators to understand if this solution can leverage its relative investments or must be revised and enriched mainly in terms of mobility control to reach higher efficiency. Mobility control of small cells adjusted according to the traffic hotspots' locations has recently gained excessive interest for the standardization of 5G networks. Actually, it is possible to develop and standardize new mechanisms and protocols where small cells follow hotspots of users with the assistance of a traffic hotspot localization algorithms [10], [11]. Controlling the mobility of small cells is supposed to ameliorate the performances of the network because the small cell will be always near the traffic hotspot which reduces the congestion in the macro cell, the handover rate and the interference to neighboring transmissions and provides better links for mobile users. Furthermore, a good mobility control of the small cell may be cost-effective since it is able to replace the deployment of many classic small cells.

*B. Manhattan mobility model*

Manhattan model is used to emulate the movement pattern of the small cell. The city is modeled by Manhattan style grid composed of horizontal and vertical streets. All streets are two-way, with one lane in each direction and small cells movements are constrained by these lanes which represent the main characteristic and the geographic restriction of this model. At an intersection of a horizontal and a vertical street, the small cell can turn left, right or go straight. In Fig. 1, the trajectory of the moving small cell is already defined. Moreover, the velocity of the small cell at a time slot is dependent on its velocity at the previous time slot. It is also restricted by the velocity of the road traffic on the same lane of the street and by the stops where the bus must take passengers. We model the mobility of the small cell in position $L_s(t) = (R_s(t), \theta_s(t))$ moving, at time $t$, with a speed of $V_s(t)$ by the following expression

$$V_s(t + dt) = min(V_{max}, V_s(t) + \beta dV_s(t)) \quad (1)$$

where $V_{max}$ is the maximum allowed velocity, $V_s(t)$ and $dV_s(t)$ are respectively the current speed and the acceleration of the small cell and $\beta \in [-1, 1]$. If $\beta$ is negative, then the moving small cell is decelerating. The small cell position at time $t + dt$ is likewise given by

$$L_s(t + dt) = L_s(t) + V_s(t)dt \quad (2)$$

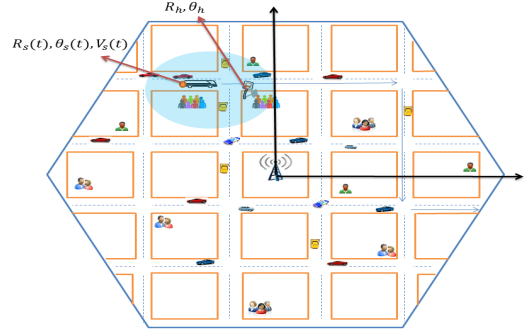

Fig. 1: A moving small cell according to Manhattan mobility model.

Note that $V_s(t)$ and $L_s(t)$ are complex numbers $\mathbb{C}$ since the small cell is moving in a bi-dimensional space.

### III. DOWNLINK SYSTEM MODEL

We consider a hexagonal cellular network with an infinite number of macro cells where the inter-site distance is denoted by $\delta$ and the transmitting power level is equal to $P$. At time $t$, a moving small cell is located at position $L_s(t)$ and offloading data traffic in its coverage area, as illustrated in Fig. 1. The transmit power level of the small cell is $P_s = \kappa P$ with $0 \leq \kappa \leq 1$.

A given UE with polar coordinates $m = (r, \theta)$, is served either by the central macro cell (cell at the origin) or the small cell, depending on the relative signal strength coming from both antennas. The rest of the neighboring macro cells play the role of interfering cells.

In order to evaluate the efficiency of moving small cells, we consider a stationary traffic hotspot with polar coordinates $(R_h, \theta_h)$. The traffic hotspot is assumed to be located inside the central macro cell, i.e. $R_h < R = \delta\sqrt{\frac{\sqrt{3}}{2\pi}}$ with $R$ is the radius of the disk having the same area as the hexagon [8]. UE locations are spatially distributed in order to form a traffic hotspot centered in $(R_h, \theta_h)$ with a standard deviation $A$; its measure is given by

$$dm(r, \theta) = \frac{1}{2\pi A^2} e^{-\frac{|re^{i\theta} - R_h e^{i\theta_h}|^2}{2A^2}} r dr d\theta \quad (3)$$

We consider two scenarios: In the first scenario, only macro cells are operational. This scenario represents a benchmark allowing the comparison of a network where small cells contribute in offloading traffic to a network without small cells. Scenario 2 considers a small cell moving near the traffic hotspot.

To model the wireless channel, we consider a distance based pathloss metric with a standard function given by

$a|m - L|^{-2b}$, where $|m - L|$ is the distance between the UE $m$ and macro or small cell position $L$ in the network. $a$ is a pathloss constant which depends on the type of the environment relative to the type of the cell (indoor, outdoor, rural, urban...) and $2b > 2$ is the pathloss exponent coefficient[1] At time $t$, the SINR received by the UE and its throughput are respectively denoted by $\gamma_t(r,\theta)$ and $\eta_t(r,\theta)$ if it is received from the macro cell and by $\tilde{\gamma}_t(r,\theta)$ and $\tilde{\eta}_t(r,\theta)$ if it is received from the small cell.

The relationship between $\gamma$ (in linear scale) and $\eta$ (in Mbps) depends on the UE capability and receiver characteristics, the available bandwidth, the radio conditions and small scale fading effects, the type of the service and the choice of the Modulation and Coding Schemes (MCS). It allows to evaluate the data rate practically experienced in any position of the network in the absence of any other active user in the cell and it is often modeled by a modified Shannon formula

$$\eta = min(K_1 \times W \times \ln(1 + K_2 \times \gamma), \eta_0) \quad (4)$$

where $K_1$ and $K_2$ are two variables depending on transmission conditions foregoing and can be adapted for each UE speed and category, $W$ is the used bandwidth and $\eta_0$ is the maximum achievable data rate of the given UE category.

## IV. PRELIMINARY ANALYSIS

We first evaluate the instantaneous user throughput obtained in each position of the covered region of the studied macro cell and small cell based on the modified Shannon formula. Then, the instantaneous throughput CCDF (Complementary Cumulative Distribution Function) is calculated considering the distribution of user locations given in (3).

The instantaneous SINRs received from the macro cell and the small cell are expressed as follows

$$\gamma_t(r,\theta) = \frac{1}{g(r) + \kappa |re^{i\theta} - R_s(t)e^{i\theta_s(t)}|^{-2b}r^{2b}} \quad (5)$$

$$\tilde{\gamma}_t(r,\theta) = \frac{P_s |re^{i\theta} - R_s(t)e^{i\theta_s(t)}|^{-2b}}{(g(r) + 1) Pr^{-2b}} \quad (6)$$

where $g(r)$ represents the interference plus noise factor in a network composed of only macro cells. It is defined by the division of the power coming from all the interfering macro cells plus the noise power by the received power from the serving macro cell.

In order to evaluate the impact of infinite number of interfering macro cells, we have established and validated in [8] an efficient and simple expression of $g(r)$. Moreover, $g$ realizes a continuous increasing function from $[0, R]$ to $[0, g(R)]$ and its inverse function is provided in [8]. The function $g$ is expressed as follows

$$g(r) = 6\alpha \left(\frac{r}{\delta}\right)^{2b} \left(\frac{1 + (1-b)^2 \left(\frac{r}{\delta}\right)^2}{(1 - \left(\frac{r}{\delta}\right)^2)^{2b-1}} + \omega(b) - 1\right) + \frac{P_N}{P}r^{2b} \quad (7)$$

where $\alpha$ is the average cell load over all the interfering macro cells, $P_N$ is the noise power level and

[1] Without loss of generality, we consider that the transmit power levels $P$ and $P_s$ include as well the pathloss constant $a$, antenna gain, cable loss, UE antenna gain and body loss.

$$\omega(b) = 3^{-b}\zeta(b)\left(\zeta(b,\frac{1}{3}) - \zeta(b,\frac{2}{3})\right)$$

$\zeta(.)$ and $\zeta(.,.)$ are respectively the Riemann Zeta and Hurwitz Riemann Zeta functions [12, pp. 1036].

In the presence of a hotspot (following a spatial distribution such as given by $dm(r,\theta)$ in (3)) in the region covered by the macro cell and the moving small cell $(R_s(t), \theta_s(t))$, we define the instantaneous throughput CCDF for both cells as follows

$$\mathbb{P}(\eta_t \geq l) = \frac{1}{S_t}\int_{S^*} \mathbb{1}\left(P_s|re^{i\theta} - R_s(t)e^{i\theta_s(t)}|^{-2b} \leq Pr^{-2b}\right) \times$$
$$\mathbb{1}\left(min(K_1 W \ln(1 + K_2 \times \gamma_t(r,\theta)), \eta_0) \geq l\right) dm(r,\theta) \quad (8)$$

$$\mathbb{P}(\tilde{\eta}_t \geq l) = \frac{1}{\tilde{S}_t}\int_{S^*} \mathbb{1}\left(P_s|re^{i\theta} - R_s(t)e^{i\theta_s(t)}|^{-2b} > Pr^{-2b}\right) \times$$
$$\mathbb{1}\left(min(K_1 W \ln(1 + K_2 \times \tilde{\gamma}_t(r,\theta)), \eta_0) \geq l\right) dm(r,\theta) \quad (9)$$

where $\mathbb{1}(.)$ is the indicator function, $S^*$ is the area covered by the macro cell and the deployed small cell and

$$S_t = \int_{S^*} \mathbb{1}\left(P_s|re^{i\theta} - R_s(t)e^{i\theta_s(t)}|^{-2b} \leq Pr^{-2b}\right) dm(r,\theta)$$

$$\tilde{S}_t = \int_{S^*} \mathbb{1}\left(P_s|re^{i\theta} - R_s(t)e^{i\theta_s(t)}|^{-2b} > Pr^{-2b}\right) dm(r,\theta)$$

It is clear that for $l > \eta_0$, the instantaneous throughput CCDFs in the macro cell and in the small cell are equal to zero since the peak throughput a UE can reach in the best radio conditions can not be higher than $\eta_0$.

$$\forall l > \eta_0, \quad \mathbb{P}(\eta_t \geq l) = 0 \quad \text{and} \quad \mathbb{P}(\tilde{\eta}_t \geq l) = 0$$

On the other hand, when $l \leq \eta_0$, the instantaneous throughput CCDFs are further simplified and are given by

$$\mathbb{P}(\eta_t \geq l) = \frac{1}{S_t}\int_{S^*} \mathbb{1}\left(P_s|re^{i\theta} - R_s(t)e^{i\theta_s(t)}|^{-2b} \leq Pr^{-2b}\right) \times$$
$$\mathbb{1}\left(\frac{1}{\gamma_t(r,\theta)} \leq \psi(l)\right) dm(r,\theta) \quad (10)$$

$$\mathbb{P}(\tilde{\eta}_t \geq l) = \frac{1}{\tilde{S}_t}\int_{S^*} \mathbb{1}\left(P_s|re^{i\theta} - R_s(t)e^{i\theta_s(t)}|^{-2b} > Pr^{-2b}\right) \times$$
$$\mathbb{1}\left(\frac{1}{\tilde{\gamma}_t(r,\theta)} \leq \psi(l)\right) dm(r,\theta) \quad (11)$$

with

$$\psi(l) = K_2 \left(e^{\frac{l}{K_1 W}} - 1\right)^{-1} \quad (12)$$

Besides, when the moving small cell leaves the coverage of the studied macro cell goes far enough from it, all UEs in the hotspot $(R_h, \theta_h)$ are served by the macro cell. Hence, the throughput CCDF can be simply given from [6] by

$$\mathbb{P}(\eta_t \geq l) = \frac{1}{S}\frac{e^{-\frac{R_h^2}{2A^2}}}{A^2}\int_0^\Lambda re^{-\frac{r^2}{2A^2}} I_0\left(\frac{rR_h}{A^2}\right) dr \quad (13)$$

with $\Lambda = min\left(g^{-1}\left(\psi(l)\right), R\right)$ and $I_0(.)$ is the first order of the modified Bessel function of the first kind. $\psi$ is defined in (12) and $g$ is given in (7).

## V. DYNAMIC LEVEL ANALYSIS OF MOVING SMALL CELLS

### A. Traffic characteristics and dynamic system description

We define the term flow, regularly used in the rest of the analysis, to refer to a session where a user successfully initiates and finishes his transmission. It is characterized by a starting time, corresponding to the time the user arrives to the system, and the size of the file to be transferred. We focus, in this work, on the case of best effort traffic where flow sizes, denoted by $\sigma$, are assumed to be mutually independent and exponentially distributed with mean $\sigma_0$ in Mbits. Flows may belong to different classes where the notion of class reflects the different clusters of users experiencing approximately the same radio conditions.

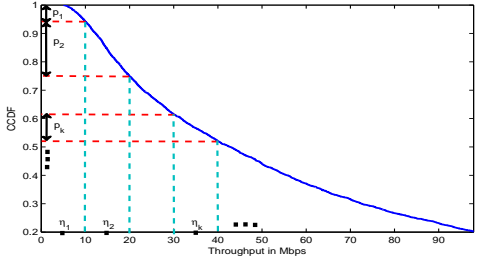

Fig. 2: Extraction of $p_{k,t}, \tilde{p}_{l,t}$ and $\eta_{k,t}, \tilde{\eta}_{l,t}$.

For tractability issues, flow classes are sorted increasingly according to the associated data rate, denoted by $\eta_{k,t}, \tilde{\eta}_{l,t}, k=1..K, l=1..L$. The probabilities $p_{k,t}, \tilde{p}_{l,t}, k=1..K, l=1..L$ denote the density of each flow class in the macro cell and the small cell respectively. Each class follows a Poisson Process of intensity $\lambda_{k,t} = \lambda_t p_{k,t} = \lambda_{Tot} \frac{S_t}{S_t+\tilde{S}_t} p_{k,t}, k=1..K$ in the macro cell and $\tilde{\lambda}_{l,t} = \tilde{\lambda}_t \tilde{p}_{l,t} = \lambda_{Tot} \frac{\tilde{S}_t}{S_t+\tilde{S}_t} \tilde{p}_{l,t}, l=1..L$ in the small cell. Data rates and flows' arrival intensities are extracted from analysis in section IV such as depicted in Fig. 2.

The service in the macro cell and the small cell are coupled. Indeed, macro cell interference on small cell users is accounted for when there is at least one active user in the macro cell. Similarly, small cell interference on macro cell users is accounted for when there is at least one active user in the small cell. It follows that $\eta_{k,t}$ and $\tilde{\eta}_{l,t}$ are given by

$$\forall k=1..K, \; \eta_{k,t} = \begin{cases} \eta_{k,t,0} & \text{if no user is served by the small cell} \\ \eta_{k,t,1} & \text{otherwise} \end{cases} \quad (14)$$

$$\forall l=1..L, \; \tilde{\eta}_{l,t} = \begin{cases} \tilde{\eta}_{l,t,0} & \text{if no user is served by the macro cell} \\ \tilde{\eta}_{l,t,1} & \text{otherwise} \end{cases} \quad (15)$$

where $\eta_{k,t,0}$ is given from the throughput distribution curve of a macro cell network in section IV and $\eta_{k,t,1}$ is given from the scenario involving a deployed moving small cell. However, for the small cell, $\tilde{\eta}_{l,t,0}$ is simply given by deleting the contribution of the central macro cell in the interference experienced by the small cell's UEs. $\tilde{\eta}_{l,t,1}$ is also derived from the throughput distribution curve obtained in section IV.

Moreover, we assume that users near the small cell[2] change from a flow class to another adjacent class after exponential durations due to the mobility of the small cell and also of the users. Handovers are also taken into account. Therefore, we assume that users handed over from a cell to another will affect the system only in the first flow class of each cell since users with bad radio conditions are the most eligible to trigger a handover and they will be highly interfered by their previous serving cell. The transition rates[3] from a flow class to another vary according to the mobility of the small cell. Hence, at time $t$, they are denoted by $\nu_{k,k+1,t}, \nu_{k,k-1,t}, \tilde{\nu}_{l,l+1,t}$ and $\tilde{\nu}_{l,l-1,t}$ (see Fig. 3). Similarly, $\nu_t$ and $\tilde{\nu}_t$ denote the instantaneous handover rates from the macro cell to the small cell and inversely. The dynamic system model is depicted in Fig. 3.

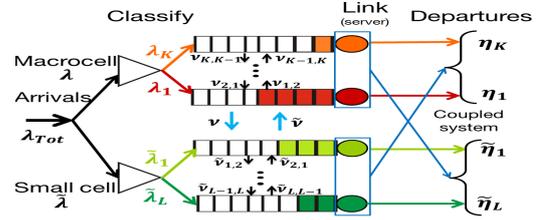

Fig. 3: Dynamic system model

### B. System coupling

We denote by $\rho$ and $\tilde{\rho}$ the load of the macro cell and the small cell respectively and by $P_0, \tilde{P}_0$ the probabilities of the steady states where no user is in the macro cell and the small cell respectively. Without considering the mobility of the small cell, for each class $k$ of flows in the macro cell, the queuing system can be modeled with a proportion $\tilde{P}_0 = 1 - \tilde{\rho}$ of flows experiencing a data rate equal to $\eta_{k,0}$ and a proportion $1 - \tilde{P}_0$ of flows experiencing a data rate equal to $\eta_{k,1}$ and same for the small cell. Hence, the system is partially modified and behaves like a network with two types of classes in both the macro cell and the small cell: macro and small cell flow classes suffering all the time from interference coming from the small cell and the macro cell respectively and flow classes not affected by this interference. Under the stability condition ($\rho, \tilde{\rho} < 1$) and following [6], the stationary distribution of the state $(n = (n_1,..,n_K), \tilde{n} = (\tilde{n}_1,..,\tilde{n}_L))$ becomes equal to

$$\pi_{n,\tilde{n}} = (1-\rho) \frac{|n|!}{\prod_{k=1}^{K} (\tilde{\rho} n_k)!((1-\tilde{\rho})n_k)!} (1-\tilde{\rho}) \frac{|\tilde{n}|!}{\prod_{l=1}^{L} (\rho \tilde{n}_l)!((1-\rho)\tilde{n}_l)!} \times$$
$$\prod_{k=1}^{K} \left(\frac{\lambda_k \sigma_0}{\eta_{k,0}}\right)^{(1-\tilde{\rho})n_k} \left(\frac{\lambda_k \sigma_0}{\eta_{k,1}}\right)^{\tilde{\rho} n_k} \prod_{l=1}^{L} \left(\frac{\tilde{\lambda}_l \sigma_0}{\tilde{\eta}_{l,0}}\right)^{(1-\rho)\tilde{n}_l} \left(\frac{\tilde{\lambda}_l \sigma_0}{\tilde{\eta}_{l,1}}\right)^{\rho \tilde{n}_l} \quad (16)$$

---

[2]i.e. in its coverage area or in the coverage area of the macro cell but experiencing a significant interfering signal coming from the small cell.

[3]The values of transition rates related to the small cell mobility are extracted from the evolution of the throughput CCDF in time.

We denote by $|n| = \sum_{k=1}^{K} n_k$ and $|\tilde{n}| = \sum_{l=1}^{L} \tilde{n}_l$ the cardinality of all the active flows in the macro cell and the small cell respectively.

*C. Performance analysis*

Analysis taking into account the mobility of the small cell is not affected by the system coupling detailed in the previous subsection since transition rates (between classes and between cells) does not depend on the queue state of the interfering cell.

The probability of having an active flow in class $k$ for the macro cell (in class $l$ for the small cell) is given by

$$\forall k = 1..K, \; P_k = \frac{E_t[n_{k,t}]}{\sum_{i=1}^{K} E_t[n_{i,t}] + \sum_{j=1}^{L} E_t[\tilde{n}_{j,t}]} \quad (17)$$

$$\forall l = 1..L, \; \tilde{P}_l = \frac{E_t[\tilde{n}_{l,t}]}{\sum_{i=1}^{K} E_t[n_{i,t}] + \sum_{j=1}^{L} E_t[\tilde{n}_{j,t}]} \quad (18)$$

where $n_t = (n_{1,t}, .., n_{K,t})$ and $\tilde{n}_t = (\tilde{n}_{1,t}, .., \tilde{n}_{L,t})$ represent the queue state in each cell and in each class at time $t$ and $E_t[h(t)] = \frac{1}{t}\sum_{s=0}^{t} h(s)$.

With the deployment of moving small cells and in the presence of traffic hotspots, the system presents two important characteristics allowing to derive several performance metrics: a high mobility and a heavy system load. Hence, following [13], the probability that a user is in class $k$ (or in class $l$) satisfies

$$\forall k = 1..K, \; q_k = \frac{E_t\left[\frac{\tilde{\nu}_t}{\nu_t} \prod_{i=1}^{k-1} \frac{\nu_{i,i+1,t}}{\nu_{i+1,i,t}} \prod_{j=1}^{L-1} \frac{\tilde{\nu}_{j,j+1,t}}{\tilde{\nu}_{j+1,j,t}}\right]}{\sum_{i=1}^{K} q_i + \sum_{j=1}^{L} \tilde{q}_j} \quad (19)$$

$$\forall l = 1..L, \; \tilde{q}_l = \frac{E_t\left[\frac{\nu_t}{\tilde{\nu}_t} \prod_{j=1}^{l-1} \frac{\tilde{\nu}_{j,j+1,t}}{\tilde{\nu}_{j+1,j,t}} \prod_{i=1}^{K-1} \frac{\nu_{i,i+1,t}}{\nu_{i+1,i,t}}\right]}{\sum_{i=1}^{K} q_i + \sum_{j=1}^{L} \tilde{q}_j} \quad (20)$$

Under the round robin policy and with considering the proposed coupling approach, it follows that the $K$ and $L$ queues are equivalent to one PS queue with service data rate given by

$$\bar{\eta} = \sum_{k=1}^{K} q_k E_t\left[\tilde{\rho}_t \eta_{k,t,1} + (1-\tilde{\rho}_t)\eta_{k,t,0}\right] +$$
$$\sum_{l=1}^{L} \tilde{q}_l E_t\left[\rho_t \tilde{\eta}_{l,t,1} + (1-\rho_t)\tilde{\eta}_{l,t,0}\right] \quad (21)$$

where $\rho_t$ and $\tilde{\rho}_t$, representing the load in both cells at time $t$, are supposed to be in $[0, 1]$, otherwise they are taken equal to 1 in the analysis. It is important to notice that this assumption affects only the coupling statement where the proportion of users with data rate $\eta_{k,t,0}$ plus those with data rate $\eta_{k,t,1}$ are equal to 100% when we have, respectively, $\tilde{\rho}_t \geq 1$ and $\rho_t \geq 1$.

The load of the cells may be higher than one since the mobility of the small cell allows to offload the extra charge from a cell to another and the system remains stable. In this context, it is proved in [13], that mobility increase the stability region in wireless networks. Then, under the stability condition, i.e. $\bar{\rho} = \frac{\lambda_{Tot}\sigma}{\bar{\eta}} < 1$, the stationary probability of state $(n = (n_1, .., n_K), \tilde{n} = (\tilde{n}_1, .., \tilde{n}_L))$ is given by

$$\pi_{n,\tilde{n}} = \frac{(1-\bar{\rho})|n+\tilde{n}|! \prod_{k=1}^{K} q_k^{n_k} \prod_{l=1}^{L} \tilde{q}_l^{\tilde{n}_l}}{\prod_{k=1}^{K} (\tilde{\rho}n_k)!((1-\tilde{\rho})n_k)! \prod_{l=1}^{L} (\rho\tilde{n}_l)!((1-\rho)\tilde{n}_l)!} \quad (22)$$

with $\rho = \lim_{t \to \infty} E_t[\rho_t]$ and $\tilde{\rho} = \lim_{t \to \infty} E_t[\tilde{\rho}_t]$.

The traffic conservation equation, which still applies, is expressed by

$$\lambda_{Tot}\sigma = \sum_{k=1}^{K} E_t\left[\eta_{k,t}\Phi_k(n_t)\right] + \sum_{l=1}^{L} E_t\left[\tilde{\eta}_{l,t}\tilde{\Phi}_l(\tilde{n}_t)\right] \quad (23)$$

where $\Phi_k(n_t) = \frac{n_{k,t}}{|n_t|}$ and $\tilde{\Phi}_l(\tilde{n}_t) = \frac{\tilde{n}_{l,t}}{|\tilde{n}_t|}$, with the round robin scheduling, represent the proportion of allocated resources to users of class $k$ and $l$ in their serving cells.

Subsequently, the mean flow throughput is given by

$$R = \sum_{k=1}^{K} P_k E_t\left[\frac{\eta_{k,t}\Phi_k(n_t)}{n_{k,t}}\right] + \sum_{l=1}^{L} \tilde{P}_l E_t\left[\frac{\tilde{\eta}_{l,t}\tilde{\Phi}_l(\tilde{n}_t)}{\tilde{n}_{l,t}}\right] \quad (24)$$

## VI. NUMERICAL RESULTS

We present in TABLE I the most important parameters used in the numerical evaluation.

TABLE I: Parameters' configuration.

| | |
|---|---|
| Macro deployment | infinite hexagonal with $\delta = 1$ Km |
| Pathloss model MtoUE | $151 + 37.6 log_{10}(d_{Km})$ |
| Pathloss model StoUE | $148 + 36.7 log_{10}(d_{Km})$ |
| BS power | Macro:46dBm, Small:30dBm |
| Antenna gain with cable loss | Macro:18dBi, Small:6dBi |
| Frequency/Bandwidth | 2.6 Ghz / 20 Mhz |
| UE category/Throughput | 3 / $\eta_0 = 98 Mbps$, $K_1 = 0.85$, $K_2 = 1.9$ |
| UE antenna gain/Body loss | 0dB / 2dB |
| File size/Scheduling/Traffic type | 2Mb / PS / FTP |

We consider a specific scenario where the position of the traffic hotspot is equal to $(R_h = 0.5, \theta_h = \frac{\pi}{3})$. We plot in Fig. 4, the throughput CCDF in a network without SCs, and also in a network with a moving SC in different times (when the SC is 0m, 60m and 120m far from the HS) and also in average (during the observation time).

In the figures' legend, MC and SC mean macro cell and small cell respectively, HS means the traffic hotspot.

From Fig. 4, we observe that the deployment of a moving small cell improves significantly the capacity of the system when it moves near the traffic hotspot since user locations with degraded radio conditions will be covered by the moving small cell and hence will experience a better SINR level. Moreover, we note that, even when the small cell moves far from the hotspot (60 meters), the system capacity is improved comparing to scenario where only macro cells are operating in the studied area. In average and during all the simulation period, the deployment of a moving small cell generates positive gains and improves the distribution of radio conditions comparing to a network without small cells.

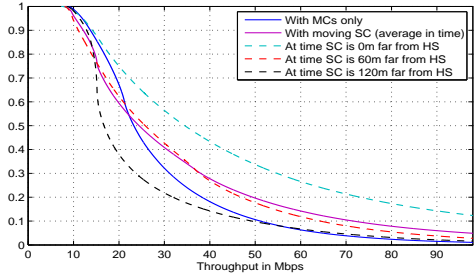

Fig. 4: Throughput CCDF defined in the preliminary study.

In Fig. 5, we plot the evolution of the load and the mean flow throughput in time for the studied scenarios: with and without a moving small cell. This later is moving according to Manhattan mobility model and is supposed to pass by the traffic hotspot periodically (after each 30 minutes) since it is located on top of buses. Curves are divided into three intervals.

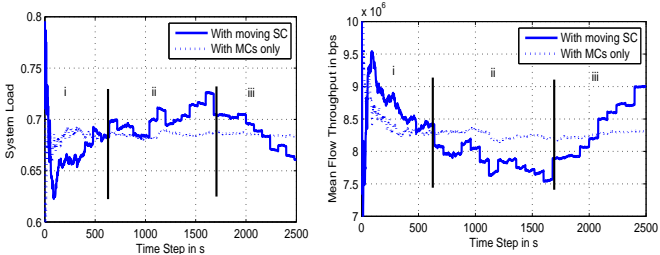

Fig. 5: The load $\bar{\rho}$ (left) and the mean flow throughput $R$ (right) in the system before and after the deployment of a moving cell.

In interval i, the load in the system is reduced with the deployment of a moving small cell and the mean flow throughput is higher than in a network with macro cells only. This is explained by the fact that the small cell is moving near the traffic hotspot located in $(0.5Km, \frac{\pi}{3})$. In interval ii, the small cell is moving far away from the traffic hotspot and higher interference (between then macro cell and the small cell) is experienced by users comparing to the first interval of time. Consequently, we observe that the load is increased with the deployment of a small cell and the mean flow throughput is reduced. This means that when the small cell is near the traffic hotspot but covering a small proportion of it, the system performances are degraded due to the significant interference on macro cell users and the increase of the proportion of cell edge users in the small cell. In interval iii, we notice that the load is reduced again and the mean flow throughput is improved. This is explained by the fact that the small cell is moving again near the traffic hotspot (since small cells are supposed to be deployed on the top of buses, the existence of small cells near the traffic hotspot is periodic which leads to the periodicity of network performances).

## VII. CONCLUSION

We studied in this paper the impact of deploying a moving small cell in the presence of stationary traffic hotspot inside a macro cell. We studied the system taking into account its dynamics where users come to the system at random times and leave it after a finite service duration, at rates taken from the preliminary analysis of this paper.

Results show that the efficiency of deploying moving small cells to offload traffic in the congested macro cell can be a beneficial solution when the small cell is moving near the traffic hotspot and covers a significant proportion of it. However, when the small cell is moving far away from the traffic hotspot, the system performance is degraded comparing to a network composed of only macro cells due to the high mutual interference between macro and small cell users.

In future work, we intend to develop a mechanism allowing to control the mobility of small cells in order to cover traffic hotspots and reduce congestion in cellular networks so as to make this promising technology a cost-effective investment.

## REFERENCES

[1] W. H. Chin, Z. Fan and R. Haines, *Emerging technologies and research challenges for 5G wireless networks*, IEEE Wireless Communications, vol. 21, no 2, p. 106-112, 2014.
[2] H. S. Dhillon, R. K. Ganti, F. Baccelli, and J. G. Andrews, *Modeling and analysis of K-tier downlink heterogeneous cellular networks*, IEEE Journal on Sel. Areas in Communications, vol. 30, no. 3, pp. 550 560, Apr. 2012.
[3] R. W. Heath, M. Kountouris, and T. Bai, *Modeling heterogeneous network interference using Poisson point processes*, IEEE Trans. on Signal Processing, vol. 61, no. 16, pp. 4114-4126, Aug. 2013.
[4] J. M. Kelif, S. Senecal and M. Coupechoux, *Impact of small cells location on performance and QoS of heterogeneous cellular networks*, in Proc. IEEE International conference PIMRC, 2013.
[5] L. Saker, S. Elayoubi, T. Chahed and A. Gati, *Energy efficiency and capacity of heterogeneous network deployment in LTE-advanced*, in European Wireless Conference, 2012. p. 1-7, 2012.
[6] A. Jaziri, R. Nasri and T. Chahed, *System level analysis of heterogeneous networks under imperfect traffic hotspot localization*, to appear in IEEE Trans. on Vehicular Technology 2016.
[7] A. Jaziri, R. Nasri and T. Chahed, *Performance Analysis of Small Cells' Deployment under Imperfect Traffic Hotspot Localization*, in IEEE Globecom 2015, Wireless Networks symposium.
[8] R. Nasri and A. Jaziri, *On the Analytical Tractability of Hexagonal Network Model with Random User Location*, to appear in IEEE Transactions on Wireless Communications 2016.
[9] Y. Sui, J. Vihriala, A. Papadogiannis, et al. *Moving cells: a promising solution to boost performance for vehicular users*, in IEEE Communications Magazine, vol. 51, no 6, p. 62-68, 2013.
[10] A. Jaziri, R. Nasri and T. Chahed, *Traffic Hotspot localization in 3G and 4G wireless networks using OMC metrics*, in Proc. IEEE PIMRC, p. 270-274, September 2014.
[11] A. Jaziri, R. Nasri and T. Chahed, *Tracking traffic peaks in mobile networks using statistics of performance metrics*, submitted.
[12] I. S. Gradshteyn and I. M. Ryzhik, *Table of integrals, series, and products*, Elsevier/Academic Press, Amsterdam, 7th edition 2007.
[13] N. Abbas, T. Bonald and B. Sayrac, *Opportunistic gains of mobility in cellular data networks*, In Proc. IEEE Modeling and Optimization in Mobile, Ad Hoc, and Wireless Networks, p. 315-322, 2015.